\documentclass{article}
\usepackage{spconf,amsmath,amssymb,amsfonts,graphicx,float,array,subcaption}
\usepackage{color}
\usepackage[colorlinks=false,linkcolor=black,anchorcolor=black,citecolor=black]{hyperref}
\usepackage{booktabs}
\usepackage[compress]{cite}

\title{RVAE-EM: Generative speech dereverberation based on recurrent variational auto-encoder and convolutive transfer function}
\name{Pengyu Wang$^{1,2}$ \qquad Xiaofei Li$^{2,*}$
\thanks{* Corresponding author}
\thanks{E-mails: wangpengyu@westlake.edu.cn, lixiaofei@westlake.edu.cn}
}
\address{
  $^1$Zhejiang University, Hangzhou, China\\
  $^2$Westlake University \& Westlake Institute for Advanced Study, Hangzhou, China}

\begin{document}
\ninept
\maketitle
\begin{abstract}
In indoor scenes, reverberation is a crucial factor in degrading the perceived quality and intelligibility of speech. 
In this work, we propose a generative dereverberation method. 
Our approach is based on a probabilistic model utilizing a recurrent variational auto-encoder (RVAE) network and the convolutive transfer function (CTF) approximation.
Different from most previous approaches, the output of our RVAE serves as the prior of the clean speech.
And our target is the maximum a posteriori (MAP) estimation of clean speech, which is achieved iteratively through the expectation maximization (EM) algorithm. 
The proposed method integrates the capabilities of network-based speech prior modelling and CTF-based observation modelling. 
Experiments on single-channel speech dereverberation show that the proposed generative method noticeably outperforms the advanced discriminative networks. 
\end{abstract}
\begin{keywords}
Speech enhancement, 
speech dereverberation, 
variational auto-encoder,
convolutive transfer function,
unsupervised learning.
\end{keywords}
\section{Introduction}
\label{sec:intro}
Speech enhancement aims to extract clean speech from degraded recordings,
which is widely adopted in applications such as 
hearing aids, online communication devices, and intelligent robots.
In indoor scenarios, the reverberation caused by reflection, refraction, and diffusion during sound wave propagation significantly degrades speech quality and intelligibility~\cite{vincent2018audio}.  
Traditional dereverberation approaches usually build models of speech and recordings, and complete the task by deconvolution.
With the advancements in deep neural networks (DNNs) and the availability of large speech datasets, 
DNN-based approaches have been proposed,
surpassing the performance of traditional ones.
DNN-based approaches can be roughly categorized into two groups based on their paradigms.

One paradigm of DNN-based approaches is to use discriminative networks to build a mapping from reverberant speech to clean ones.
The mapping can be implemented in terms of waveforms~\cite{su2019perceptually} or spectrograms~\cite{zhao2020monaural,xiong2022spectro}.
For example, TCN-SA~\cite{zhao2020monaural} and FullSubNet~\cite{hao2020fullsubnet,zhou2023speech} map the degraded spectrogram to clean ones,
and achieve the state-of-the-art (SOTA) dereverberation performance.
The discriminative networks are usually trained in a supervised manner with paired clean speech and corresponding reverberant recordings.
One limitation of discriminative networks is the difficulty of covering all possible acoustic conditions in the training dataset.
Sometimes, these methods may not perform as well in unseen acoustic environments~\cite{richter2022speech}.

A different paradigm is to regard speech dereverberation as a generative task
and use generative networks such as 
variational auto-encoders (VAEs)~\cite{Kingma2014auto},
generative adversarial networks (GANs)~\cite{goodfellow2020generative}, 
and diffusion models (DMs)~\cite{ho2020denoising}
to complete it.
For example, in \cite{baby2021speech}, the authors combined the VAE network with non-negative matrix factorization (NMF) to realize unsupervised speech dereverberation, 
which resulted in better performance than the traditional NMF-based approaches.
In \cite{su2020hifi}, the authors combined the end-to-end WaveNet with adversarial training to tackle noise, reverberation, and distortion.
In \cite{richter2022speech}, the authors developed a DM-based iterative speech enhancement approach, yielding remarkable performance in both denoising and dereverberation.

In this work, we propose a generative dereverberation method named RVAE-EM.
Instead of studying the direct mapping from reverberant speech to clean ones, our objective is the maximum a posteriori (MAP) estimation of clean speech.
To achieve this, we build a probabilistic model, which describes the process of generating observed recordings.
In the short-time Fourier transform (STFT) domain,
we train a recurrent variational auto-encoder (RVAE)~\cite{leglaive2020recurrent} to model the generative process of clean speech,
and use the convolutive transfer function (CTF) approximation~\cite{talmon2009relative} to model the observation process. 
The expectation maximization (EM) algorithm~\cite{moon1996expectation} is used to iteratively estimate the clean spectrogram and the acoustic parameters in the observation process.
Our method integrates the capabilities of DNN-based speech prior modelling and CTF-based observation modelling. 
The proposed RVAE network can be trained in both unsupervised (U) and supervised (S) manners, resulting in two versions: RVAE-EM-U and RVAE-EM-S.
Experiments on single-channel dereverberation show that RVAE-EM-U achieves SOTA performance in the unsupervised manner, while RVAE-EM-S noticeably outperforms the advanced discriminative networks~\cite{zhao2020monaural,zhou2023speech}.


\section{Models}
Our approach is based on a probabilistic model, which describes the process of generating observed recordings from latent variables.
The probabilistic model consists of two parts: a generative model of clean speech and an observation model of reverberant recordings.

\subsection{Clean speech generative model based on RVAE network}
The RVAE is a dynamic VAE network that consists of an encoder (parameterized by $\boldsymbol{\phi}$) and a decoder (parameterized by $\boldsymbol{\theta}$). 
Following the usage of RVAE for speech denoising in \cite{bie2022unsupervised}, we train an RVAE to obtain the prior of clean speech in the STFT domain.

Let $S_{f}(n)\in \mathbb{C}^{1}$ denote the clean spectrogram (dry speech spectrogram) in a specific time-frequency (T-F) bin, 
where $n\in\left[1,N\right]$ is the frame index and $f\in\left[1,F\right]$ is the frequency index.
We suppose that the clean spectrogram $\mathbf{S}\in\mathbb{C}^{F\times N}$ follows a zero-mean complex Gaussian distribution, and this distribution 
is generated according to standard Gaussian distributed latent variables $\mathbf{Z}\in\mathbb{R}^{D\times N}$, $D \ll F$. 
The prior probability of latent variables is frame-independent as
\begin{equation}
    \label{pz}
    p\left(\mathbf{Z}\right)
    = \textstyle\prod_{n=1}^N
    p\left(\mathbf{Z}(n)\right)
    = \textstyle\prod_{n=1}^N
    \mathcal{N}\left(\mathbf{Z}(n);\mathbf{0},\mathbf{I}_{D}\right),
\end{equation}
where $\mathbf{Z}(n) \in \mathbb{R}^{D \times 1}$ is the $n$th column of $\mathbf{Z}$, 
denoting the latent variables at the $n$th frame.
The conditional generative process of clean spectrogram is modelled by the RVAE decoder as
\begin{equation}
    \label{ps|z}
    \begin{aligned}
        p_{\boldsymbol{\theta}}\left(\mathbf{S}|\mathbf{Z}\right)
        &=\textstyle\prod_{n=1}^N
        p_{\boldsymbol{\theta}}\left(\mathbf{S}(n)|\mathbf{Z}\right) \\
        &=\textstyle\prod_{n=1}^N
        \mathcal{CN}\left(\mathbf{S}(n);\mathbf{0},\boldsymbol{\Sigma}_{\mathrm{spch}}(n)\right),
    \end{aligned}
\end{equation}
where $\mathbf{S}(n) \in \mathbb{C}^{F \times 1}$ is the $n$th column of $\mathbf{S}$, denoting the clean spectrogram at the $n$th frame, 
and $\boldsymbol{\Sigma}_{\mathrm{spch}}(n) = \mathrm{diag}([\sigma^2_{\mathrm{spch},1}(n),$
$\sigma^2_{\mathrm{spch},2}(n),\cdots,\sigma^2_{\mathrm{spch},F}(n)])$ 
is a diagonal matrix whose diagonal entries are the output of RVAE decoder at frame $n$.
In each frame, 
the acquisition of $\boldsymbol{\Sigma}_{\mathrm{spch}}(n)$
takes into account latent variables from all frames.
The generative model of clean spectrogram is
\begin{equation}
    p_{\boldsymbol{\theta}}\left(\mathbf{S},\mathbf{Z}\right)=p_{\boldsymbol{\theta}}\left(\mathbf{S}|\mathbf{Z}\right)p\left(\mathbf{Z}\right).
\end{equation}


To train the RVAE decoder, the encoder must be introduced.
The encoder provides a parameterized Gaussian posterior 
\begin{equation}
    \label{qz|s}
    \begin{aligned}
    q_{\boldsymbol{\phi}}\left(\mathbf{Z}|\mathbf{S}\right)
    = \textstyle\prod_{n=1}^N
    q_{\boldsymbol{\phi}}\left(\mathbf{Z}(n)|\mathbf{Z}(1:n-1),\mathbf{S}\right),
    \end{aligned}
\end{equation}
which is an approximation of the intractable posterior $p_{\boldsymbol{\theta}}\left(\mathbf{Z}|\mathbf{S}\right)$.
In each frame, the posterior of $\mathbf{Z}(n)$ is obtained according to 
both the latent variables from all previous frames
and the clean spectrogram from all frames.
The word 'recurrent' in RVAE describes the time dependency between clean spectrogram $\mathbf{S}$ and latent variables $\mathbf{Z}$.

The RVAE is trained by jointly optimizing the encoder and decoder.
The details of our RVAE will be shown in Section \ref{sec:RVAE}.


\subsection{Observation model based on CTF approximation}
The room impulse response (RIR) is the propagating filter in an enclosure.
The CTF approximation is commonly utilized to model the reverberation effect in the STFT domain since the RIR length is relatively longer than the STFT window. 
In this work, we follow the CTF formulation described in \cite{li2017algorithm}. 

Considering the case of a single speaker and a single microphone, 
assuming the CTF length is $P+1$ and $P \ll N$,
the observed recording can be modelled as
\begin{equation}
\label{x2}
    \begin{aligned}
        X_{f}(n)
        \approx \textstyle\sum_{p=0}^P
        S_{f}(n-p)H_{f}(p)
        + W_{f}(n),
    \end{aligned}
\end{equation}
where $X_{f}(n)\in \mathbb{C}^{1}$ and $W_{f}(n)\in \mathbb{C}^{1}$ are the STFT coefficients of reverberant speech and additive noise, respectively. 
$H_{f}(p)\in \mathbb{C}^{1}$ is the CTF filter coefficient.
We can rewrite it in the matrix form as
\begin{equation}
\label{x3}
    \begin{aligned}
        \mathbf{X}_{f}
        \approx \widetilde{\mathbf{H}}_{f}\mathbf{S}_f
        +\mathbf{W}_{f},
    \end{aligned}
\end{equation}
where $\mathbf{X}_{f} = \left[X_{f}\left(1\right),X_{f}\left(2\right),\cdots,X_{f}\left(N\right)\right]^T\in \mathbb{C}^{N \times 1}$, 
$\mathbf{S}_{f} = \left[S_{f}\left(1\right),S_{f}\left(2\right),\cdots,S_{f}\left(N\right)\right]^T\in \mathbb{C}^{N \times 1}$,
\\
\begin{equation}
\begin{aligned}
\widetilde{\mathbf{H}}_{f} = 
\begin{bmatrix}
  &H_{f}(0)  &0  &\cdots &\cdots  &0 \\
  &\vdots   &\ddots &0  &\ddots   &\vdots  \\
  &H_{f}(P)  &\ddots &H_{f}(0)  &\ddots  &\vdots  \\
  &0  &\ddots &\vdots  &\ddots  &\vdots \\
  &\vdots &\ddots  &H_{f}(P)&\ddots   &0 \\
  &0  &\cdots &0  &\cdots &H_{f}(0)
\end{bmatrix} 
\in \mathbb{C}^{N \times N},
\end{aligned}
\end{equation}
and $\mathbf{W}_{f} = \left[W_{f}\left(1\right),W_{f}\left(2\right),\cdots,W_{f}\left(N\right)\right]^T\in \mathbb{C}^{N \times 1}$.

Supposing that the noise is isotropic stationary complex Gaussian, there is
\begin{equation}
    p_{\boldsymbol{\psi}}\left(\mathbf{X}_f|\mathbf{S}_f\right) = \mathcal{CN}\left(\mathbf{X}_f;\widetilde{\mathbf{H}}_{f}\mathbf{S}_f,\boldsymbol{\Sigma}_{\mathrm{noi},f}\right),
\end{equation}
where $\boldsymbol{\Sigma}_{\mathrm{noi},f}=\sigma_{\mathrm{noi},f}^2\mathbf{I}_N$, $\sigma_{\mathrm{noi},f}^2$ and $\mathbf{I}_N$ are the noise power in the $f$th band and $N \times N$ identity matrix, respectively.  
$\boldsymbol{\psi}$ is the set of acoustic parameters which consists of the CTF filter coefficients and noise power.

Defining the full-band observed recording as
$\mathbf{X} \in \mathbb{C}^{F\times N}$, we have the observation model
\begin{equation}
    p_{\boldsymbol{\psi}}\left(\mathbf{X}|\mathbf{S}\right)
    =\textstyle\prod_{f=1}^{F}p_{\boldsymbol{\psi}}\left(\mathbf{X}_f|\mathbf{S}_f\right).
\end{equation}

\begin{figure*}[thbp]
  \centering
    \begin{minipage}[b]{0.40\textwidth}
        \centering
        \includegraphics[width=0.95\textwidth]{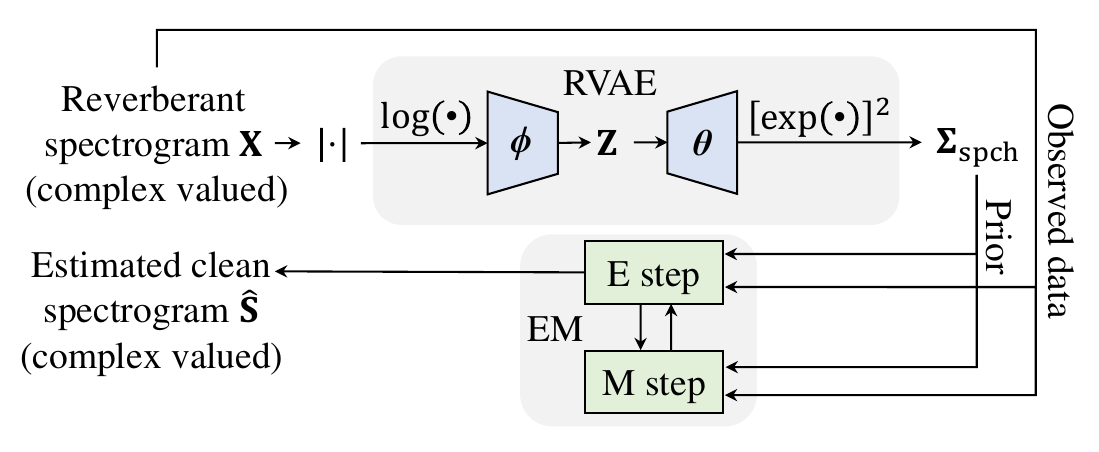}
        \vspace{10pt}
        \subcaption{Overview.}
        \label{fig:overview}
    \end{minipage}
    \begin{minipage}[b]{0.48\textwidth}
        \centering
        \includegraphics[width=0.95\textwidth]{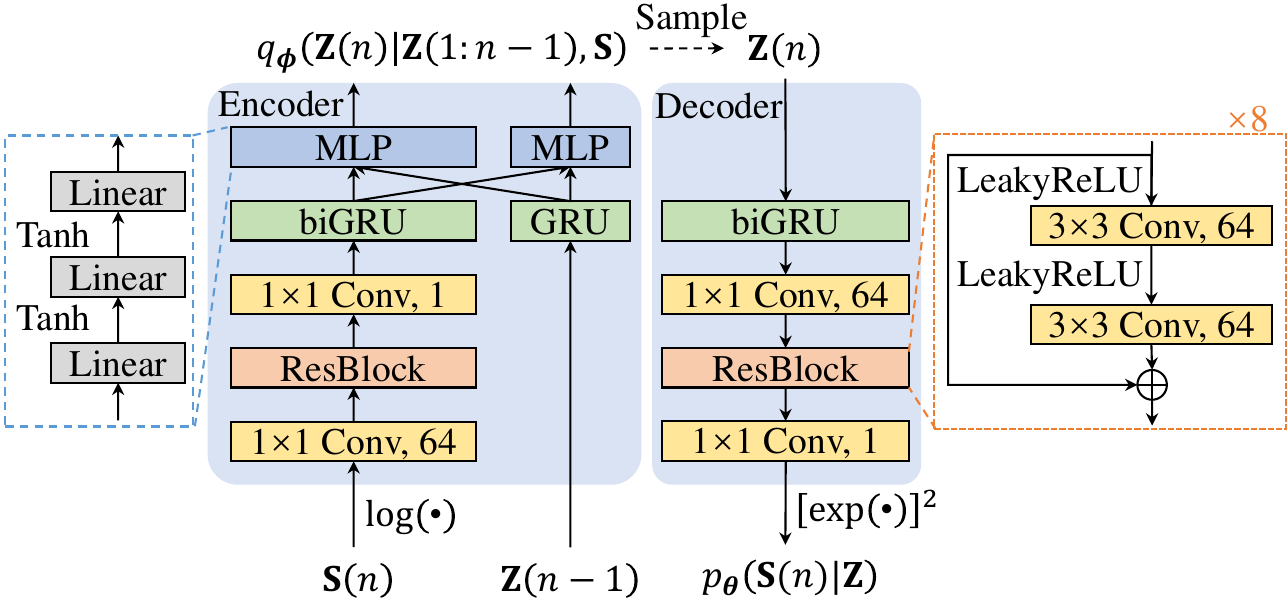}
        \vspace{-5pt}
        \subcaption{RVAE architecture.}
        \label{fig:RVAE}
    \end{minipage}
    \vspace{-10pt}
    \caption{Overview of our method and the proposed RVAE architecture.}
  \vspace{-10pt}
\end{figure*}

\section{Proposed method}
\label{sec:3}
In our method, 
we use the EM algorithm to estimate the clean spectrogram and the acoustic parameters iteratively,
which has two stages: E step and M step.

\subsection{E step: Estimation of clean spectrogram}
The E step provides a MAP estimate of the clean spectrogram given the corresponding reverberant input.
The objective is expressed as
\begin{equation}
    \label{Shat}
    \begin{aligned}
        \widehat{\mathbf{S}}
        = \mathrm{E}_{p\left(\mathbf{S}|\mathbf{X}\right)}\left[\mathbf{S}\right] 
        = \mathrm{E}_{p\left(\mathbf{Z}|\mathbf{X}\right)}\left[\mathrm{E}_{p\left(\mathbf{S}|\mathbf{Z},\mathbf{X}\right)}\left[\mathbf{S}\right]\right]. 
    \end{aligned}
\end{equation}

For the inner expectation of (\ref{Shat}), according to the Bayes rules and our probabilistic assumptions, we have
\begin{equation}
    \begin{aligned}
        p\left(\mathbf{S}|\mathbf{Z},\mathbf{X}\right)
        = \frac{
            p\left(\mathbf{X}|\mathbf{S},\mathbf{Z}\right)
            p\left(\mathbf{S}|\mathbf{Z}\right)
            }
            {p\left(\mathbf{X}|\mathbf{Z}\right)} 
        \propto
            p_{\boldsymbol{\psi}}\left(\mathbf{X}|\mathbf{S}\right)
            p_{\boldsymbol{\theta}}\left(\mathbf{S}|\mathbf{Z}\right).
    \end{aligned}
\end{equation}
It can be found that $p\left(\mathbf{S}|\mathbf{Z},\mathbf{X}\right)$ is also Gaussian: 
\begin{equation}
    \begin{aligned}
        p\left(\mathbf{S}|\mathbf{Z},\mathbf{X}\right)
        =\textstyle\prod_{f=1}^{F}
        \mathcal{CN}\left(\mathbf{S}_f;\boldsymbol{\mu}_f,\boldsymbol{\Sigma}_{f}\right),
    \end{aligned}
\end{equation}
where
\begin{equation}
\begin{aligned}
\label{eq:posterior}
    \boldsymbol{\Sigma}_{f}
    =\left(\widetilde{\mathbf{H}}_f^H\boldsymbol{\Sigma}_{\mathrm{noi},f}^{-1}\widetilde{\mathbf{H}}_f+\boldsymbol{\Sigma}_{\mathrm{spch},f}^{-1}\right)^{-1},
\end{aligned}
\end{equation}
\begin{equation}
\begin{aligned}
    \boldsymbol{\mu}_{f}
    =\boldsymbol{\Sigma}_{f}\widetilde{\mathbf{H}}_f^H\boldsymbol{\Sigma}_{\mathrm{noi},f}^{-1}\mathbf{X}_f,
\end{aligned}
\end{equation}
and $\mathbf{\Sigma}_{\mathrm{spch},f}= \mathrm{diag}([\sigma^2_{\mathrm{spch},f}(1),\sigma^2_{\mathrm{spch},f}(2),\cdots,\sigma^2_{\mathrm{spch},f}(N)])$ is a diagonal matrix whose diagonal entries are the output of RVAE decoder in the $f$th band.

However, there are still two issues:
both the posterior $p\left(\mathbf{Z}|\mathbf{X}\right)$ and the outer expectation of (\ref{Shat}) are intractable. 
We approximate $p\left(\mathbf{Z}|\mathbf{X}\right)$ with $q_{\boldsymbol{\phi}}\left(\mathbf{Z}|\mathbf{X}\right)$, and sample the latent variables $\mathbf{Z}_\mathrm{sample}$.
Using the sampled latent variables to approximate the expectation, the clean spectrogram is estimated as
\begin{equation}
    \widehat{\mathbf{S}}
    \approx \mathrm{E}_{p\left(\mathbf{S}|\mathbf{Z}_{\mathrm{sample}},\mathbf{X}\right)}\left[\mathbf{S}\right].
\end{equation}
This means, in practice, the RVAE inference is performed once to obtain the prior variance of clean speech, i.e. $\boldsymbol{\Sigma}_{\mathrm{spch},f}$, which will be used for all E-steps (and all M-steps), and the clean speech estimation is set to $\widehat{\mathbf{S}}_{f} \approx  \boldsymbol{\mu}_{f}$ for each $f$.

\subsection{M step: Estimation of acoustic parameters}

After the E step, the acoustic parameters in the observation model can be updated by maximizing the expected logarithmic likelihood.

For the CTF parameters 
${\mathbf{H}}_f = [H_f(0),H_f(1),\cdots,H_f(P)] \in \mathbb{C}^{1 \times (P+1)}$
in each band, we have
\begin{equation}
\begin{aligned}
    \widehat{\mathbf{H}}_f 
    &= \mathop{\arg\max}\limits_{\mathbf{H}_f}
    \mathrm{E}_{p\left(\mathbf{S},\mathbf{Z}|\mathbf{X}\right)}
    \left[\ln{p\left(\mathbf{X},\mathbf{S},\mathbf{Z}\right)}\right] \\
    &= \left(\textstyle\sum_{n=0}^N X_f\left(n\right)\boldsymbol{\mu}^H_f\left(n:n-P\right)\right) \left(\textstyle\sum_{n=0}^N\widetilde{\boldsymbol{\Sigma}}_f\left(n\right)\right)^{-1},
\end{aligned}
\end{equation}
where
\begin{equation}
\begin{aligned}
    \widetilde{\boldsymbol{\Sigma}}_f\left(n\right) 
    & = \boldsymbol{\mu}_f\left(n:n-P\right)\boldsymbol{\mu}_f^H\left(n:n-P\right) \\
    & + \boldsymbol{\Sigma}_{f}\left(n:n-P,n:n-P\right).
\end{aligned} 
\end{equation}

Similarly, the noise power in each band is updated as
\begin{equation}
\begin{aligned}
    \widehat{\sigma}_{\mathrm{noi},f}^2 
    &= \mathop{\arg\max}\limits_{{\sigma}_{\mathrm{noi},f}^2}
    \mathrm{E}_{p\left(\mathbf{S},\mathbf{Z}|\mathbf{X}\right)}
    \left[\ln{p\left(\mathbf{X},\mathbf{S},\mathbf{Z}\right)}\right] \\
    &= \frac{{\lVert\mathbf{X}_f-\widetilde{\mathbf{H}}_f\boldsymbol{\mu}_f\rVert_F^2}
    +\mathrm{tr}\left(\widetilde{\mathbf{H}}_f\boldsymbol{\Sigma}_f\widetilde{\mathbf{H}}_f^H\right)}
    {N},
\end{aligned}
\end{equation}
where $\lVert \cdot \rVert_F$ is the Frobenius norm and $\mathrm{tr}\left(\cdot\right)$ is the trace of a matrix.

After the M step, we can proceed to the E-step to further update the clean spectrogram.
An overview of our method is shown in Fig.~\ref{fig:overview}. 
The RVAE network provides the prior of clean speech. 
And the EM algorithm achieves the MAP estimation of the clean spectrogram according to both prior knowledge and observed recordings.

\subsection{Details of RVAE}
\label{sec:RVAE}

We propose a new network architecture for RVAE, which is more powerful for speech representation learning compared to other RVAE networks~\cite{bie2022unsupervised}. 
We set the feature dim of latent variables as $D=32$.
%
The specific architecture of the proposed RVAE is shown in Fig.~\ref{fig:RVAE}.
\emph{Conv}s 
are convolutional 
layers.
\emph{ResBlock}s consist of eight modules in series. 
Each module consists of the LeakyReLU activation function, two $3\times3$ \emph{Conv}s, and skip connections. 
The numbers of in-channels and out-channels of \emph{ResBlock} are both 64.
\emph{GRU} and \emph{biGRU}s are single-layer forward and bidirectional gate recurrent units with 256 and 512 hidden units, respectively.
\emph{MLP}s consists of fully-connected layers with 256, 256, and 32 units. 
The Tanh activation is used after all layers in \emph{MLP}s except the last layer.
Dropout of $20\%$ is applied in \emph{ResBlock}s
and the first two layers of \emph{MLP}s.
To transform the feature scales, operation $\mathrm{log}(\cdot)$ is applied before the encoder.
And operation $\left[\mathrm{exp}(\cdot)\right]^2$ is applied after the decoder to get the prior variance of the clean spectrogram.

In our network, \emph{Conv}s in the encoder are used for spectral feature extraction.
After that, the features are sent to \emph{biGRU} for temporal feature extraction.
The hidden state of \emph{biGRU} contains information of clean spectrogram for all frames.
On the other hand, 
at each frame, the input and hidden state of \emph{GRU} keeps the information of all previous latent variables.
Therefore, the outputs of \emph{biGRU} and \emph{GRU} contains the information of $\mathbf{S}$ and $\mathbf{Z}(1:n-1)$ respectively.
Then, the outputs are sent to two \emph{MLP}s to get the mean and variance of $q_{\boldsymbol{\phi}}\left(\mathbf{Z}(n)|\mathbf{Z}(1:n-1),\mathbf{S}\right)$ separately.
Between the encoder and decoder, the reparameterization trick~\cite{Kingma2014auto} is utilized to sample the latent variables from the given distribution.
Similarly, the output of \emph{biGRU} in the decoder contains information of $\mathbf{Z}$.
And it is further sent to \emph{Conv}s to get the distribution $p_{\boldsymbol{\theta}}\left(\mathbf{S}(n)|\mathbf{Z}\right)$.
Since this distribution is zero-mean complex Gaussian, 
the final output in each frame is its variance $\boldsymbol{\Sigma}_{\mathrm{spch}}(n)$.

The training of RVAE is achieved by maximizing the evidence lower bound.
For simplicity, given an $N$-frame input, we define $\boldsymbol{\Sigma}_{\mathrm{spch}}\in\mathbb{R}^{F\times N}$ as the output of RVAE composed of $\sigma^2_{\mathrm{spch},f}(n)$ for each $f\in [1,F]$ and $n \in [1,N]$.
The training loss is
\begin{equation}
    \label{loss}
    \begin{aligned}
        \mathcal{L}
        &=-\mathrm{E}_{q_{\boldsymbol{\phi}}\left(\mathbf{Z}|\mathbf{S}\right)}\left[\ln p_{\boldsymbol {\theta}}\left(\mathbf{S}|\mathbf{Z}\right)\right]
        +\mathrm{D}_{KL}\left(q_{\boldsymbol{\phi}}\left(\mathbf{Z}|\mathbf{S}\right)||p\left(\mathbf{Z}\right)\right) \\
        &= \mathrm{D}_{IS}\left({|\mathbf{S}|}^2,\boldsymbol{\Sigma}_{\mathrm{spch}}\right) + \mathrm{D}_{KL}\left(q_{\boldsymbol{\phi}}\left(\mathbf{Z}|\mathbf{S}\right)||p\left(\mathbf{Z}\right)\right),
    \end{aligned}
\end{equation}
where
$\mathrm{D}_{IS}\left(\cdot\right)$ denotes the Itakura-Saito (IS) divergence~\cite{itakura1968analysis}  
and 
$\mathrm{D}_{KL}\left(\cdot\right)$ denotes the Kullback-Leibler (KL) divergence~\cite{kullback1951information}.
Because only clean speech is utilized during training,
the whole method becomes unsupervised.
We denote this version as RVAE-EM-U.

After the unsupervised training,
we can also fine-tune the RVAE with reverberant-clean-paired training data to get a more accurate posterior distribution of latent variables.
The fine-tuning loss is
\begin{equation}
    \label{lossft}
    \begin{aligned}
        \mathcal{L}_{\mathrm{ft}} 
        &=-\mathrm{E}_{q_{{\boldsymbol{\phi}}}\left(\mathbf{Z}|\mathbf{X}\right)}\left[\ln p_{\boldsymbol {\theta}}\left(\mathbf{S}|\mathbf{Z}\right)\right]
        +\mathrm{D}_{KL}\left(q_{{\boldsymbol{\phi}}}\left(\mathbf{Z}|\mathbf{X}\right)||p\left(\mathbf{Z}\right)\right) \\
        &= \mathrm{D}_{IS}\left({|\mathbf{S}|}^2,\boldsymbol{\Sigma}_{\mathrm{spch}}\right) + 
        \mathrm{D}_{KL}\left(q_{{\boldsymbol{\phi}}}\left(\mathbf{Z}|\mathbf{X}\right)||p\left(\mathbf{Z}\right)\right).
    \end{aligned}
\end{equation}
This supervised version is denoted as RVAE-EM-S.

\begin{table*}[htb]
\caption{Results of dereverberation on simulated datasets.}
\vspace{-10pt}
\label{tab:result1}
\centering
\resizebox{0.99\linewidth}{!}{
\begin{tabular}{cccccccccccc}
\toprule
Method & Type & Params & SISDR$\uparrow$ & WBPESQ$\uparrow$ & NBPESQ$\uparrow$ & STOI$\uparrow$ & ESTOI$\uparrow$ & SRMR$\uparrow$ & MOS-SIG$\uparrow$ & MOS-BAK$\uparrow$ & MOS-OVRL$\uparrow$\\
\midrule
Unprocessed
&-
&-
&-7.33 ± 5.44
&1.25 ± 0.16
&1.59 ± 0.23
&0.69 ± 0.10
&0.45 ± 0.14
&3.38 ± 1.41
&2.20 ± 0.69
&2.11 ± 0.66
&1.69 ± 0.46\\
\cmidrule(lr){0-11}
VAE-NMF~\cite{baby2021speech}
& G\&U
&7.5M
&-6.45 ± 5.73
&1.36 ± 0.25
&1.74 ± 0.31
&0.74 ± 0.09
&0.52 ± 0.13
&4.34 ± 1.68
&2.66 ± 0.54
&2.65 ± 0.63
&2.05 ± 0.48
\\
\textbf{RVAE-EM-U} (prop.) 
& G\&U 
&7.0M
&-4.96 ± 6.16
&1.62 ± 0.38
&2.07 ± 0.42
&0.81 ± 0.08
&0.64 ± 0.13
&6.37 ± 2.75
&3.06 ± 0.38
&2.96 ± 0.63
&2.39 ± 0.47\\
\cmidrule(lr){0-11}
TCN-SA~\cite{zhao2020monaural}
&D\&S
&4.7M
&-4.10 ± 5.20
&2.27 ± 0.36
&2.68 ± 0.34
&0.92 ± 0.03
&0.81 ± 0.05
&7.50 ± 2.73
&3.14 ± 0.23
&3.78 ± 0.22
&2.80 ± 0.26
\\
FullSubNet~\cite{hao2020fullsubnet}
&D\&S
&14.5M
&-3.99 ± 5.22
&2.39 ± 0.38
&2.78 ± 0.37
&0.92 ± 0.03
&0.81 ± 0.05
&6.69 ± 2.14
&3.00 ± 0.28
&3.67 ± 0.28
&2.64 ± 0.31
\\
SGMSE+~\cite{richter2022speech}
&G\&S
&65.6M
&7.40 ± 4.50
&2.61 ± 0.46
&3.04 ± 0.40
&0.96 ± 0.02
&0.88 ± 0.05
&7.99 ± 4.32
&3.56 ± 0.10
&4.02 ± 0.14
&3.26 ± 0.15\\
RVAE (w/o EM) (prop.)
&G\&S
&7.0M
&-4.21 ± 4.95
&1.97 ± 0.28
&2.39 ± 0.27
&0.89 ± 0.03
&0.75 ± 0.06
&6.43 ± 2.35
&3.09 ± 0.26
&3.49 ± 0.39
&2.64 ± 0.33
\\
\textbf{RVAE-EM-S} (prop.) 
& G\&S
&7.0M
&-1.55 ± 5.40
&2.49 ± 0.34
&2.95 ± 0.32
&0.93 ± 0.03
&0.84 ± 0.05
&8.92 ± 3.78
&3.29 ± 0.21
&3.75 ± 0.35
&2.92 ± 0.30  \\
\bottomrule
\end{tabular}}
\vspace{-10pt}
\end{table*}

\section{Experiments}

\subsection{Settings}

\textbf{Datasets}
Parts of the speaker-independent clean speech sampled at 16kHz from the Wall Street Journal (WSJ0) corpus~\cite{paul1992design} are used for training, validating and testing.
The training and test datasets consist of 24.92 and 1.48 hours of recordings, respectively.
We simulate reverberant-dry RIR pairs with gpuRIR toolbox~\cite{diaz2021gpurir} for fine-tuning and testing.
The length, width, and height of simulated rooms are randomly selected in the range $\left[5\mathrm{m},15\mathrm{m}\right]$, $\left[5\mathrm{m},15\mathrm{m}\right]$, and $\left[2\mathrm{m},6\mathrm{m}\right]$, respectively. 
The minimum distance from the speaker/microphone to the wall is $1\mathrm{m}$.
Similar to the datasets created in \cite{richter2022speech},
for reverberant RIRs, the reverberation time RT60s are uniformly chosen in the range $\left[0.4\mathrm{s},1\mathrm{s}\right]$.
Correspondingly, dry RIRs are generated by using the same geometric parameters but an absorption coefficient of 0.99.
The input and target speech signals are generated by convoluting the WSJ0 utterances with reverberant and dry RIRs, respectively.  
The training and test datasets consist of 5000 and 300 RIRs respectively.

\textbf{Other settings}
The input and output of our method are complex-valued one-sided spectrograms 
using the Hanning window with a length of 1024 samples and a hop length of 256 samples.
Additionally, we set the direct-current component of the spectrograms to zero and dismiss it.
Therefore, the input and output have $F=512$ frequency bands.
The utterances are divided into segments of 5.104 seconds (320 frames) for training and testing.

We set the CTF length in the observation model to $P=30$.
The initialization of the EM algorithm is always important.
We set the iteration steps to 100 
and initialize $\sigma_{\mathrm{noi},f}^2=10^3\times{\left|\mathbf{X}_f\right|}^2/N$, 
$H_f\left(0\right)=1$ and $H_f\left(\neq 0\right)=0$ for each $f \in [1,F]$.

During training, 
the batch size is 64.
We use the AdamW optimizer~\cite{loshchilov2017decoupled} with a maximum learning rate of 0.0001.
Particularly, we warm up the KL loss periodically to mitigate KL vanishing in RVAE~\cite{fu_cyclical_2019_naacl}.
Codes and speech examples are available on \href{https://github.com/Audio-WestlakeU/RVAE-EM}{https://github.com/Audio-WestlakeU/RVAE-EM}.

\textbf{Comparison methods}
The comparison methods include 
VAE-NMF~\cite{baby2021speech}, 
TCN-SA~\cite{zhao2020monaural},
FullSubNet~\cite{hao2020fullsubnet}, 
and 
SGMSE+~\cite{richter2022speech}.
These approaches cover generative (G) and discriminative (D) models, 
as well as supervised (S) and unsupervised (U) training manners.
To demonstrate the role of the EM algorithm,
we also show the results of our fine-tuned RVAE without any EM step.
Notice that in TCN-SA we use the reverberant phase without enhancement. 
For FullSubNet, we adopt the modified version described in \cite{zhou2023speech}, 
which slightly changes the network architecture and training target.

\textbf{Evaluation metrics}
The evaluation metrics include SISDR~\cite{Vincent2006Performance}, 
WBPESQ~\cite{rix2001perceptual},
NBPESQ~\cite{rix2001perceptual}, 
STOI~\cite{taal2011algorithm}, 
ESTOI~\cite{jensen2016algorithm}, 
SRMR~\cite{falk2010non}
and MOS-SIG, MOS-BAK, MOS-OVRL from DNSMOS~\cite{reddy2021dnsmos}.
Higher metrics denote better speech quality and intelligibility.

\subsection{Results and discussion}
Table \ref{tab:result1} shows the results of single-channel dereverberation on the simulated datasets.
In the unsupervised manner, our RVAE-EM-U surpasses the previous SOTA approach, VAE-NMF.
That is mainly because the proposed RVAE architecture is much more powerful than the MLP-based VAE used in VAE-NMF.
As a result, our network can provide a more credible prior of clean speech for the subsequent EM algorithm.
Because we train the RVAE with dry input but test it with reverberant input, the prior provided by RVAE may not be ideal.
However,
we can still estimate the clean speech to an extent, 
since the MAP estimation depends on not only the prior knowledge but also the observed data.

All supervised approaches exhibit superior performance to unsupervised ones due to the provision of paired training data.
Our RVAE-EM-S outperforms the two SOTA discriminative networks, i.e. TCN-SA and FullSubNet, which shows the capability and potential of the proposed (first-try of) generative-observation model. The diffusion-model-based SGMSE+ achieves better performance than RVAE-EM-S.
However, the proposed RVAE network has much fewer parameters. In addition, although RVAE-EM-S and SGMSE+ are both iterative during inference, SGMSE+ performs one network inference for each iteration, while the RVAE network in our approach does not participate in the EM iterations, which further saves time and memory cost.
The performance of RVAE (w/o EM) is not as good as that of RVAE-EM-S
as it solely utilizes the RVAE-based generative model and disregards the CTF-based observation model.
In fact,
the EM algorithm can reconstruct the phase and revise the magnitude of the spectrogram estimated by RVAE.

\begin{figure}
    \centering
    \includegraphics[width=0.95\columnwidth]{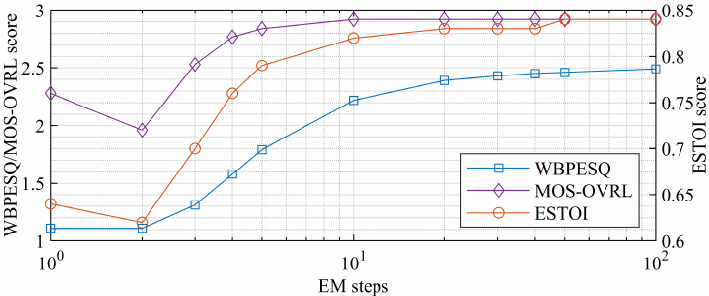}
    \vspace{-10pt}
    \caption{Average metrics of RVAE-EM-S versus EM steps.}
    \label{fig:metrics}
    \vspace{-13pt}
\end{figure}


In Fig.~\ref{fig:metrics}, we plot the average WBPESQ, ESTOI, and MOS-OVRL of RVAE-EM-S as the number of EM steps increases.
The performance measures first decrease at the second step, possibly because the initialized CTF coefficients are overly simplified, and then a degraded clean speech will be estimated in the E-step. However, this problem will be corrected after updating the CTF coefficients in the first M-step.   
As the number of iteration steps increases, the performance of RVAE-EM-S rapidly improves up to the first 10 steps, and then slowly converges. 
It demonstrates that, through maximizing the likelihood of observations, the EM iterations are consistently improving the estimation of clean speech and acoustic parameters.   

\section{Conclusions}
In this work, we propose a speech dereverberation method named RVAE-EM.
Our work is built upon a probabilistic model that comprises a generation model for clean speech, which is based on the RVAE network, and an observation model, which is based on the CTF approximation.
The output of our method provides a MAP estimate of the clean spectrogram, which integrates the information from both clean speech generative model and observation model.
Our method has two versions, RVAE-EM-U and RVAE-EM-S, which are trained in the unsupervised and supervised manner, respectively. 
Experiments on single-channel dereverberation demonstrate that RVAE-EM-U achieves SOTA performance among unsupervised methods, 
while RVAE-EM-S significantly surpasses the majority of existing supervised approaches.


\clearpage
\bibliographystyle{IEEEbib}
\bibliography{refs}

\end{document}